\documentclass[twocolumn,prl,preprintnumbers,amsmath,amssymb]{revtex4}

\newcommand{\be}
{\begin{eqnarray}}

\newcommand{\ee}
{\end{eqnarray}}

\usepackage{graphicx}
\usepackage{dcolumn}
\usepackage{bm}
\usepackage{times}

\usepackage{color}
\definecolor{rot}{rgb}{1,0,0}
\definecolor{blau}{rgb}{0,0,1}

\begin{document}

\bibliographystyle{unsrt}
\title{Quantum-dot-spin single-photon interface}
\author{S. T. Y\i lmaz$^{*}$\footnote[0]{\scriptsize$^{\rm *}$ These authors have contributed equally to this work.}}
\author{P. Fallahi$^{*}$}
\author{A. Imamo\u{g}lu}
\affiliation{Institute of Quantum Electronics, ETH Zurich, CH-8093
Z\"{u}rich, Switzerland}
\date{\today}

\begin{abstract}

Using background-free detection of spin-state-dependent resonance
fluorescence from a single-electron charged quantum dot with an
efficiency of $0.1\%$, we realize a single spin-photon interface
where the detection of a scattered photon with 300 picosecond time
resolution projects the quantum dot spin to a definite spin
eigenstate with fidelity exceeding $99\%$. The bunching of
resonantly scattered photons reveals information about electron spin
dynamics. High-fidelity fast spin-state initialization heralded by a
single photon enables the realization of quantum information
processing tasks such as non-deterministic distant spin
entanglement. Given that we could suppress the measurement
back-action to well below the natural spin-flip rate, realization of
a quantum non-demolition measurement of a single spin could be
achieved by increasing the fluorescence collection efficiency by a
factor exceeding $20$ using a photonic nanostructure.
\end{abstract}

\maketitle

Optical excitations of an electron spin confined in self-assembled
quantum dots (QD) have favorable selection rules that allow for
recycling trion transitions where the scattered photon polarization
is strongly correlated with the electron spin-state
\cite{Imamoglu:1999}. Realization of a spin-photon interface in the
spirit of what has been recently realized for trapped ions
\cite{Monroe:2004} however, suffers from the fact that the
background excitation laser scattering from the solid-state
interfaces and defects overwhelms the QD resonance fluorescence.
While single QD resonance fluorescence has been recently observed by
several groups \cite{Vamivakas:2009,Flagg:2009}, the reported
experiments did not demonstrate a complete suppression of the
background in a charge-controlled QD that is essential for the
realization of a spin-photon interface.

In this Letter, we demonstrate background-free detection of single
QD resonance fluorescence (RF) with an efficiency of $0.1 \%$. We
show that detection of a single photon, resonantly scattered on the
charged-exciton (trion) resonance, projects the QD spin to a state
where the spin is pointing along the external magnetic field
($|\uparrow\rangle$ ) with a conditional initialization fidelity of
$99.2 \%$. Our results constitute a first step towards the
realization of non-deterministic spin-photon \cite{Monroe:2004} and
spin-spin entanglement \cite{Cabrillo:1999,Monroe:2007} schemes.
Given that we operate in a regime where measurement back-action in
the form of spin-flip raman scattering rate is weaker than the
natural spin-flip rate induced by exchange coupling to a fermionic
reservoir, our results also constitute a first step towards quantum
non-demolition (QND) measurement \cite{Walls-Milburn} of a single
solid-state spin. We estimate that either by identifying QDs with a
small heavy-light-hole mixing \cite{Fernandez:2009} or by embedding
a QD in a two dimensional photonic crystal structure
\cite{Finley:2008} or a micro-cavity \cite{Hennessy:2007}, it will
be possible to realize a QND measurement.

The experiments are carried out with self-assembled InGaAs QDs
embedded in a Shottky diode heterostructure which enables
deterministic loading of electrons from a nearby Fermi sea separated
by a 35 nm thick blocking barrier. The sample is studied using a
diffraction limited confocal microscope housed in a liquid-helium
bath cryostat at 4.2 K. A single QD is addressed using a laser that
is tuned into resonance with its fundamental exciton or trion
resonance. An external magnetic field $B$ is applied in the Faraday
configuration. Absorption measurements are performed by detecting
the transmitted photons using a photo-diode placed underneath the
sample. Scattered photons are collected through an optical fiber and
detected using an avalanche photo-diode (APD). A zirconium solid
immersion lens (SIL) is used to improve the QD collection
efficiency.

\begin{figure}[t]
\includegraphics[scale=1]{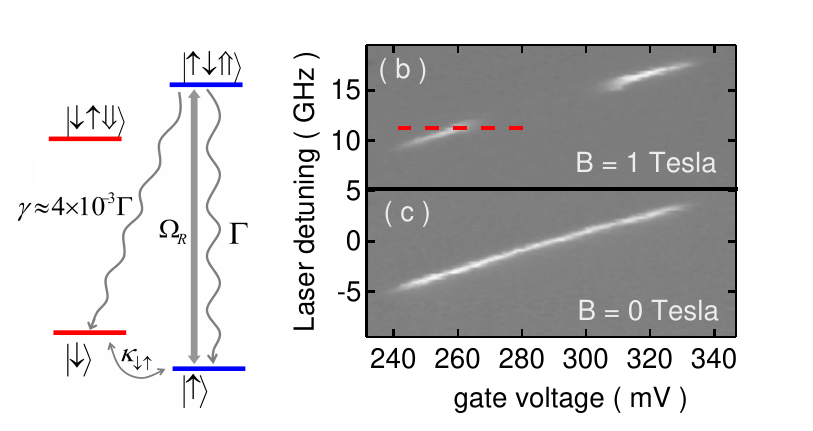}
\caption{\label{Fig1} (a) Energy level diagram for a quantum dot
(QD) charged with a single electron. (b)\&(c) Differential
Transmission (DT) signal as a function of gate voltage and laser
frequency at $B=1T$ and $B=0T$. At $B=0T$ DT signal is seen at gate
voltages where the QD is singly charged. At $B=1T$ the DT signal
(white points) in the middle of the plateau disappears due to spin
pumping. Dashed line corresponds to the gate voltage trace in
Fig.~\ref{Fig2}(a)}
\end{figure}
Figure~\ref{Fig1}(a) illustrates the energy levels relevant for the
experiment. In the ground state a single electron resides in the
dot, either in the spin up $|\uparrow\rangle$ or spin down
$|\downarrow\rangle$ eigenstate. The two optically excited (trion)
states $|\uparrow\downarrow\Uparrow\rangle$ and
$|\uparrow\downarrow\Downarrow\rangle$
are formed by two electrons in
the singlet state and a heavy hole, either in the (pseudo) spin up
$|\Uparrow\rangle$ or down $|\Downarrow\rangle$ state. The two
ground and excited states are split by the Zeeman energies $|g_{e}|
\mu _{B}B$ and $|g_{h}| \mu _{B}B$. The optical selection rules in
the Faraday geometry are such that a right (left)-hand circularly
polarized optical field drives the transition $|\uparrow\rangle$
$\leftrightarrow$ $|\uparrow\downarrow\Uparrow\rangle$
($|\downarrow\rangle$ $\leftrightarrow$
$|\downarrow\uparrow\Downarrow\rangle$ ) with the Rabi frequency
$\Omega_{R}$ ($\Omega_{L}$). Spontaneous emission from the trion
state occurs with rate $\Gamma \sim 10^{9} s^{-1}$ and is circularly
polarized. The selection rules are relaxed by the ground state
mixing due to hyperfine interaction of the electron with the nuclei
and by the heavy-light-hole mixing, resulting in weak diagonal
transitions with rate {$\gamma$}. For the experiments reported here
at $B = 1T$ the branching ratio $\frac{\gamma}{\gamma+\Gamma}$ is
measured to be $\sim 4 \times 10^{-3}$ and is most likely determined
by a small in-plane (external) magnetic field component mixing the
electron spin states. Due to the interaction with the electrons in
the Fermi sea, electron spin flip events occur with a co-tunneling
rate $\kappa_{\uparrow\downarrow}$; this rate can be tuned by more
than 6 orders of magnitude from $10^7 s^{-1}$ to about $10^1 s^{-1}$
by changing the gate voltage \cite{Dreiser:2008}.

To determine the transition energies of the QD differential
transmission (DT) measurements are carried out, where the
transmitted field is recorded with a photo-diode under the sample.
In the experiments presented here, we have only addressed the higher
energy (blue) trion transition $|\uparrow\rangle$ $\leftrightarrow$
$|\uparrow\downarrow\Uparrow\rangle$ with a resonant laser field;
provided that $B> 0.2$T, excitation of the $|\downarrow\rangle$
$\leftrightarrow$ $|\downarrow\uparrow\Downarrow\rangle$ transition
can be safely neglected. Fig.~\ref{Fig1}(b)\&(c) show the DT signal
from the trion transition with a 100 ms integration time at $B=1T$
and $B=0T$ . The linear increase in the transition energy with
increasing gate voltage is due to DC Stark effect. In the middle of
the single electron plateau the DT signal disappears at $B=1T$ since
in this gate voltage regime the rate of the laser induced spin
pumping  into $|\downarrow\rangle$ state far exceeds the
co-tunneling rate $\kappa_{\uparrow\downarrow}$, resulting in high
fidelity preparation of the electron in the $|\downarrow\rangle$
state within several microseconds \cite{Atature:2006}. At the
plateau edges, high co-tunneling rate ensures the randomization of
the electron spin population. The broadening of the absorption
lineshape for gate voltages where the DT signal starts to disappear
is due to dynamical nuclear spin polarization \cite{Latta:2009}.

\begin{figure}
\centering
\includegraphics[scale=1.1]{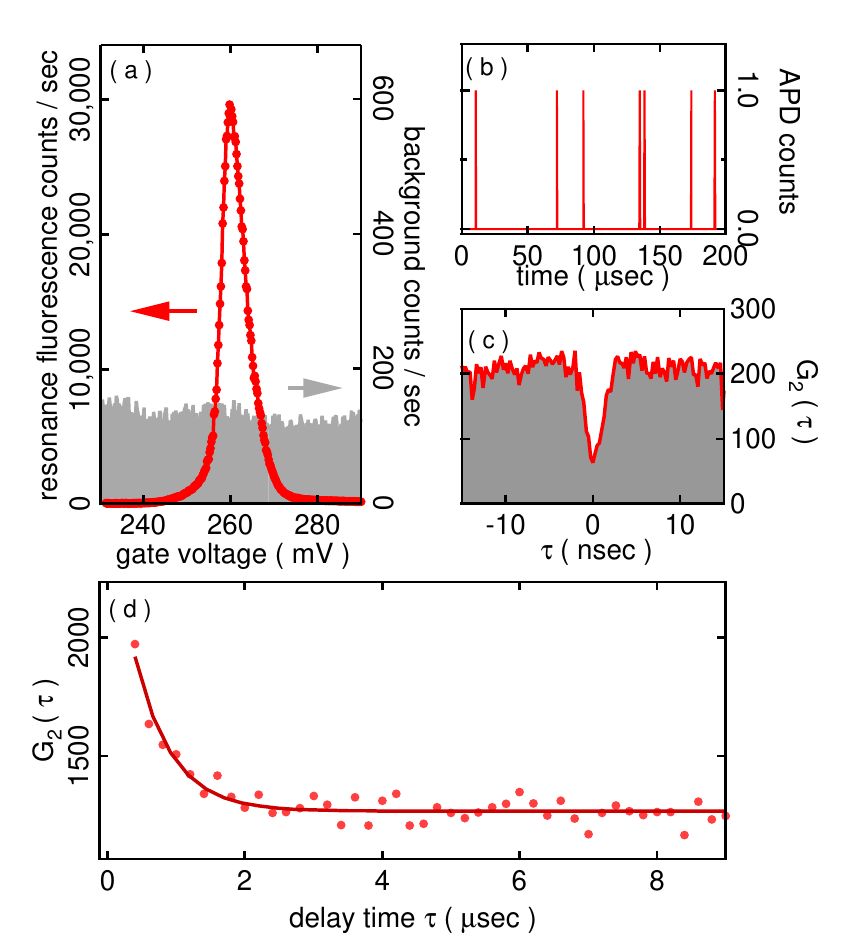}
\centering \caption{\label{Fig2} (a) Resonance fluorescence (RF)
signal from the QD as the gate voltage is scanned along the red
dashed line on Fig.~\ref{Fig1}b. Laser power is well below the QD
saturation, $P = 0.1\cdot P_{sat}$. The gray trace shows the
measurement background obtained at the same laser power with the
laser frequency fully detuned from the QD transition. On resonance
the ratio of RF photons to laser background exceeds $200$. (b) A
typical time trace recorded from the avalanche photo-diode (APD)
with a 200 nsec time resolution with a resonant laser with $P
=0.1\cdot P_{sat}$. Each pulse arises from the detection of a single
photon, which indicates that the spin is in the $|\uparrow\rangle$
state with $(99.2\pm 0.1)\%$ fidelity. (c) $G_{2}$ curve obtained by
measuring photon coincidences on two APDs on nsec timescales. The
expected antibunching behavior for a single emitter is observed,
with a spontaneous emission rate $\Gamma \sim 10^{9} s^{-1}$. The
$G_{2}$ curve does not reach zero at $\tau=0$ due to the finite time
resolution $\sim 450\,psec$ of the Hanbury-Brown and Twiss
measurement set-up. (d) Unnormalized photon correlation, $G_{2}$
curve obtained from $\sim 60,000$ traces such as the trace in (b)
for $P=0.1\cdot P_{sat}$. Solid line is an exponential fit with a
decay time $\tau_{decay}=(540 \pm 40)$nsec, corresponding to the
cotunneling limited lifetime of the $|\uparrow\rangle$ state.}
\end{figure}

RF is detected by collecting the emitted QD photons through the
focussing objective (NA = 0.55) and a single mode fiber to an APD in
the Geiger mode.  A linear polarizer that is placed before the
collection fiber and oriented orthogonal to the reflected laser
polarization extinguishes the reflected laser photons by a factor
exceeding $10^{6}$ while eliminating only half of the circularly
polarized RF photons. This high level of polarization extinction was
possible by ensuring that the deviations from linear polarization in
the excitation and collection paths are minimized.
Fig.~\ref{Fig2}(a) shows the RF signal as the gate voltage is
scanned along the dashed line in Fig~1(b) with an excitation laser
power 10 times below the QD saturation. The deviation of the RF
lineshape from the expected Lorentzian is most likely due to the
nuclear spin polarization \cite{Latta:2009}, since we observe that
for $B=0$, the RF lineshape has a perfect Lorentzian lineshape. At
the gate voltage when the laser is on resonance with the
$|\uparrow\rangle$ $\leftrightarrow$
$|\downarrow\uparrow\Uparrow\rangle$ transition, the ratio of the RF
counts to the total photon counts is $0.996\pm0.001$. Taking into account the branching ratio, the fidelity of our conditional state initialization is $0.992\pm0.001$. Based on measured RF
counts ($\sim 500,000$/sec) at $B = 0T$ for a laser power above QD
saturation and using the measured trion lifetime of $\sim 1$nsec, we
estimate an unprecedented overall QD RF collection efficiency of
$\sim 0.1\%$.

Fig.~\ref{Fig2}(b) shows a typical time trace of the APD counts
integrated for 200 ns per point as the gate voltage in
Fig.~\ref{Fig2}(a) is kept constant for resonant excitation. Whereas
at a given time the absence of a photon detection provides almost no
information about the electron spin-state, detection of an APD
electrical pulse means with a confidence level of 99.2 $\%$ that the
electron is in the $|\uparrow\rangle$ state. Moreover, we can locate
the electron spin projection event onto the $|\uparrow\rangle$ state
associated with each photon count to within $300$psec of the arrival
of the APD pulse; this projection time uncertainty is limited only
by the jitter in the rise-time of the APD pulse.

We study RF count statistics by performing second order correlation
($G_{2}$) analysis. We first note that the RF photon correlations at
short time-intervals exhibit the hallmark photon antibunching
signature of a single quantum emitter (Fig.~\ref{Fig2}(c)).
Fig.~\ref{Fig2}(d) shows the unnormalized photon correlation
$G_{2}(\tau)$ curve obtained from $\sim 60,000$ time traces such as
the one shown in Fig.~\ref{Fig2}(b). Since laser photons follow
Poissonian statistics with a flat $G_{2}(\tau)$ curve, the bunching
behavior around $\tau=0$ is a signature of RF photons, revealing
information about the spin-flip dynamics. The decay of
$G_{2}(\tau)$, fitted to an exponential function with the decay time
$\tau_{decay}=(540 \pm 40)$ $ns$, provides a direct measurement of
the $|\uparrow\rangle$ lifetime. As we discuss below, for the laser
power used in this experiment, the laser induced spin decay rate is
$(4.4  \mu s)^{-1}$ and the spin lifetime is almost exclusively
determined by co-tunneling processes. Given our collection
efficiency and the measured spin lifetime, we estimate that the
probability that we detect a photon while the electron is in
$|\uparrow\rangle$ state is $\sim (2.0\pm0.2) \%$. By varying the laser
intensity the efficiency of spin-state detection can be improved to
about $10 \%$; this enhancement comes at the expense of an increase
in the laser background such that only $98.6\%$ of the detected
photons originate from the QD RF.

\begin{figure}
\includegraphics[scale=1]{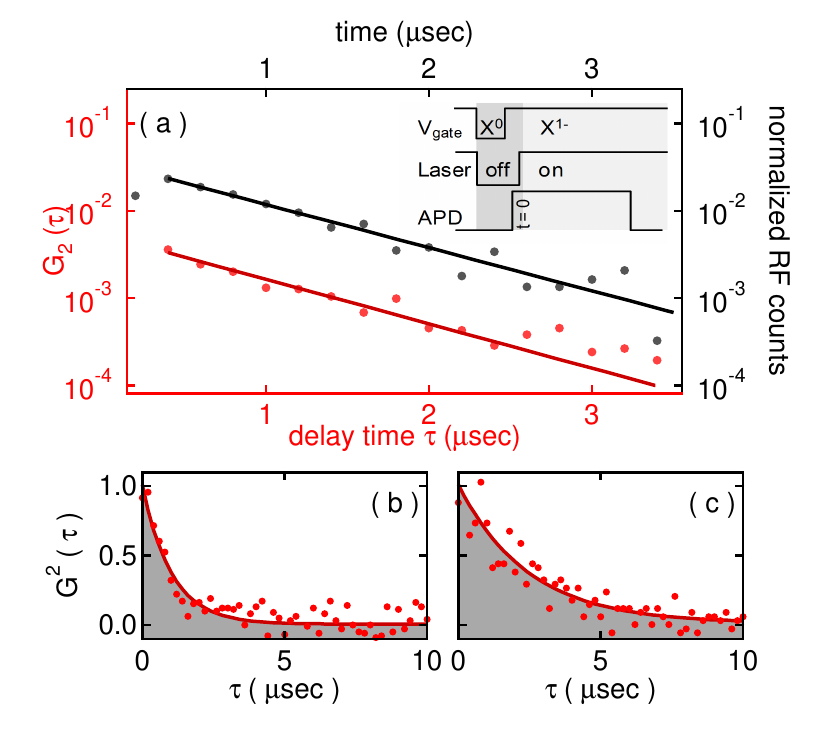}
\centering \caption{\label{Fig3}(a) $G_{2}$ (red) and nshot (black)
measurements obtained at a gate voltage in the middle of the plateau and laser intensity
$P=P_{sat}$. Solid lines are exponential fits with decay times $(840\pm40)ns$ and
$(860\pm20)ns$ respectively, demonstrating the agreement between the two measurement
methods. (b)\&(c) $G_{2}$ curves obtained from a second quantum dot with a laser
power $P=0.1\cdot P_{sat}$ and at two gate voltages close to the plateau
edge (large cotunneling) and $7.5 mV$ apart. The solid lines are
exponential fits, showing decay times of $(1.0\pm0.1)\mu s$ and
$(2.5\pm0.2)\mu s$ respectively. These decay times are direct measurement
of cotunneling rates at the given gate voltages, with the faster decay,
(b), corresponding to the gate voltage closer to the plateau edge as
expected.}
\end{figure}

In order to prove that the $G_{2}(\tau)$ decay time is indeed a
measure of the spin lifetime, we perform an n-shot measurement
\cite{Lu:2010} of the laser induced spin decay time in the middle of
the single electron plateau where the co-tunneling rate is
negligible and compare the result to the $G_{2}(\tau)$ measurements.
The scheme of an n-shot measurement cycle is depicted in the inset
of Fig.~\ref{Fig3}(a). In these n-shot spin measurements, a linearly
polarized laser on resonance with the $|\uparrow\rangle$
$\leftrightarrow$ $|\uparrow\downarrow\Uparrow\rangle$ transition is
switched on and over a time window of micro seconds the RF counts are
saved with 200 $ns$ integration time per data point. To undo
spin-pumping that arises from the spin-measurement back-action in
the form of spontaneous spin-flip Raman scattering into the
$|\downarrow\rangle$ state, we apply a gate voltage pulse that first
ejects the electron from the QD and then injects a new electron with
a random spin. Repeating this cycle $\sim 10,000$ times at a laser
power of $P_{sat}=3 nW$ corresponding to QD saturation , we obtain
the exponentially decaying n-shot curve depicted in
Fig.~\ref{Fig3}(a) (black points) with the decay time of
$(860\pm20)ns$. The decay time of $G_{2}(\tau)$ obtained with the
same laser power and at the same gate voltage is $(840\pm40)ns$ and
is in good agreement with the n-shot measurement at the
corresponding laser power (Fig.~\ref{Fig3}(a) (red points)). The
laser induced decay rates are expected to be proportional to the
trion population, which in turn is proportional to the RF counts.
The measured decay times of $(480\pm30)ns$, $(860\pm20)ns$ and
$(4.4\pm0.2)\mu s$ from the n-shot measurements at laser powers $10
P_{sat}$, $P_{sat}$ and $0.1\cdot P_{sat}$ agree well with the
expected ratio of $1:2:10$ (data not shown). Fig.~\ref{Fig3} (b),(c)
show $G_{2}(\tau)$ measurements from another QD performed at two
different gate voltages at the edge of the single electron plateau.
In this regime the spin life time is determined by the cotunneling
processes and the decay of the curves reveals the dependence of the
cotunneling rate on the gate voltage.

The realization of a spin-photon interface constitutes a key step
towards  implementation of quantum information processing protocols
such as non-deterministic spin entanglement between distant spins.
Even though elimination of laser background through polarization
suppression in our scheme results in the loss of correlations
between the spin-state and the emitted photon polarization, the fact
that the photons emitted by the two spin-states have different
energies ensures that by driving both the red and the blue trion
transitions resonantly with a $\pi$-pulse, we could generate the
entangled spin-photon state $|\psi \rangle =
(|\uparrow,1_{blue}\rangle - |\downarrow,1_{red}\rangle)/\sqrt{2}$,
starting from an initial electron spin in state $(|\uparrow \rangle
- |\downarrow \rangle)/\sqrt{2}$. As was demonstrated by Monroe and
co-workers \cite{Monroe:2007} for two trapped ions, such entangled
spin-photon states generated from two distant QDs could be used to
achieve spin entanglement conditioned upon coincidence detection of
one blue-trion and one red-trion photon at the output of a
Hong-Ou-Mandel interferometer. We emphasize that in the
non-deterministic entanglement experiments using trapped ions, the
entanglement fidelity was primarily limited by the detector dark
counts that were $\sim 20\%$ of the signal photons: these
considerations highlight the significance of the factor of $20$
improvement in the fluorescence to background photon ratio we have
achieved in our experiments, as compared to prior work
\cite{Lu:2010}. The principal challenge in realization of the
distant QD spin-entanglement scheme is the identification of two QDs
with similar enough trion resonances. Even if the two QD trion
resonance energies are not exactly identical, it is possible to
generate identical QD photons by off-resonant Rayleigh scattering
\footnote [1]{In the case of two non-identical QDs, it would be
advantageous to use the Voigt geometry, where one starts from an
initial electron spin state $|\uparrow_x \rangle$ and upon
scattering of a laser photon project the system onto the state
$|\phi \rangle = (|\uparrow_x,1_{blue}\rangle +
|\downarrow_x,1_{red}\rangle)/\sqrt{2}$. In this case, the energy
difference between the blue and the red photons is determined
exclusively by the electron Zeeman energy.}. On the other hand,
nearly identical QD pairs that can be tuned onto resonance using the
gate-voltage-induced dc-Stark shift have already been identified in
other experiments \cite{Finley:personal}. More importantly,
demonstration of two-photon interference of single-photon pulses
generated by two different QDs has also been demonstrated
\cite{Flagg:2010}. Locking of trion resonances to the resonant laser
field via dynamical nuclear spin polarization could be used to
ensure that the electron Zeeman splitting of the two QDs are
rendered identical \cite{Latta:2009}.

The background-free spin-photon interface  that we have realized
could be considered as a non-deterministic method for ultra-fast
conditional spin-state initialization with high fidelity. The
deterministic methods for spin-state preparation rely on optical
pumping where a fidelity exceeding $99\%$ could only be achieved on
a timescale $\sim 10 \mu$s for Faraday \cite{Atature:2006} and
$\sim10 n$s for Voigt geometry \cite{Xu:2007}. In contrast,
detection of a single resonance fluorescence photon in our case
prepares the spin state with the same level of initialization
fidelity on a timescale limited only by the detector response time
\footnote[2]{We emphasize that the average waiting time for
achieving this single-photon-detection-based initialization is still
$\sim 10 \mu$s for the Faraday geometry that we used.}. For the
APD's we have used, this timescale is $300 $ps; however with faster
single-photon detectors, it would be straightforward to achieve a
timescale of $40$ ps. Given the ultra-short $T_2^* \sim 2$nsec
characteristic of electron spins in self-assembled QDs \cite{Dreiser:2008}, fast
spin-state initialization is important for protocols relying on
preparation of the spin in a superposition state. We also predict
that fast spin initialization could be useful in carrying out
conditional electron spin resonance (ESR) measurements \cite{Koppens:2006}
without the need for a pulsed microwave source;
we envision  here that the detection of a photon initializes the
electron spin in the spin-up state, which then undergoes Rabi
oscillations under the influence of the continuous-wave resonant
microwave field. The likelihood of detecting a second photon at a
time $\tau$ later will oscillate with this Rabi coupling, such that
photon coincidences at time delay $\tau$ will reveal information
about coherent spin rotation.

From a quantum measurement perspective, our experiments realize a
positive operator valued measure (POVM) with measurement operators
$\hat{E}_1 = p_1 |\uparrow\rangle \langle\uparrow|$ and $\hat{E}_2 =
|\downarrow\rangle\langle\downarrow| + (1-p_1)
|\uparrow\rangle\langle\uparrow|$ \cite{Nielsen-Chuang}. For the
parameters of Fig. 2, the probability $p_1 \simeq 0.02$ for a
measurement time of $\sim0.5 \mu$s, but can be increased to $0.1$ as mentioned before. If the collection efficiency and
the excitation laser power were increased such that $p_1 \sim 1.0$
was achieved, our scheme would constitute a single-shot QND
measurement of the electron spin. We estimate that such a spin
measurement \cite{Hanson:2007} is within reach using our scheme. Collection
efficiencies from microcavities have been predicted to be as high as
$35\%$ \cite{Larque:2009}. One possibility is to use QDs coupled to
gated photonic crystal microcavities that are engineered for high
collection efficiency \cite{Englund:2009,Strauf:2007}. We predict
that such structures could give up to a factor of 15 improvement in
the overall collection efficiency. In addition, given the
observation that a sizeable fraction of QDs have vanishing in-plane
heavy-hole g-factors \cite{Fernandez:2009}, it is plausible to expect much smaller
spin-flip spontaneous emission rate from the trion state $\gamma$,
as compared to what we observed in our experiments. A smaller
branching ratio will reduce the laser back action allowing a
non-demolition measurement to be performed on a longer time scale,
and hence with better efficiency. By combining the above two
factors, achieving a factor of more than $20$ improvement in the
measurement efficiency $p_1$ necessary for a single-shot QND
measurement appears feasible.

A particularly exciting possible extension for the spin-photon
interface demonstrated here is for a double-QD system consisting of
a gate-defined and a self-assembled QD \cite{Engel:2006}. It has
been shown that in a coupled-QD system, optical transitions in a
neutral QD could be used to monitor the spin-state of a
single-electron charged QD \cite{Kim:2008}. If
realized, a double QD spin-photon interface along the lines we
describe in this Letter could be used to generate distant spin
entanglement between two gated QDs. Another interesting extension is
the implementation of our method in a single-hole charged QD where
the positively-charged trion transition selection rules are
identical to the ones we considered \cite{Gerardot:2008}; because of
the larger $T_2^* $-time of the hole spin, distant hole-spin
entanglement protocol would be easier to verify experimentally.

We would like to acknowledge the contributions of Thomas Volz and Gemma Fernandez during
the early stages of this work. We thank Ajit Srivastava for his help
with the $G_{2}$ measurement.  We also thank to Antonio Badolato for
the sample growth and Mete Atature for helpful discussions. S.T. Y\i lmaz acknowledges financial
support from the European Union within the Marie-Curie
Training Research Network EMALI. This work is supported
by NCCR Quantum Photonics (NCCR QP), research
instrument of the Swiss National Science
Foundation (SNSF).


\begin{thebibliography}{}
\bibitem{Imamoglu:1999}
A.\ Imamoglu \textit{et al.}, Phys.\ Rev.\ Lett.\ \textbf{83},
4204 (1999).

\bibitem {Monroe:2004}
B.\ B.\ Blinov \textit{et al.}, Nature (London) \textbf{428}, 153
(2004).

\bibitem {Vamivakas:2009}
A.\ N.\ Vamivakas \textit{et al.}, Nature Phys. \textbf{5}, 198
(2009).

\bibitem {Flagg:2009}
E.\ B.\ Flagg \textit{et al.}, Nature Phys. \textbf{5}, 203
(2009).

\bibitem {Cabrillo:1999}
C.\ Cabrillo \textit{et al.}, Phys.\ Rev.\ A\ \textbf{59},
1025 (1999).

\bibitem {Monroe:2007}
D.\ L.\ Moehring \textit{et al.}, Nature (London) \textbf{449}, 68
(2007).

\bibitem{Walls-Milburn}
D.\ F.\ Walls and G.\ J.\ Milburn, \textit{Quantum Optics} (Springer, 2008), 2nd edition.

\bibitem{Fernandez:2009}
G.\ Fernandez \textit{et al.}, Phys.\ Rev.\ Lett.\ \textbf{103},
087406 (2009).

\bibitem{Finley:2008}
M.\ Kaniber \textit{et al.}, Phys.\ Rev.\ B\ \textbf{77},
073312 (2008).

\bibitem{Hennessy:2007} K.\ Hennessy \textit{et al.}, Nature (London) \textbf{445}, 896
(2007).

\bibitem{Dreiser:2008}
J.\ Dreiser \textit{et al.}, Phys.\ Rev.\ B\ \textbf{77},
075317 (2008).

\bibitem{Atature:2006}
M.\ Atature \textit{et al.}, Science \textbf{312},
551 (2006).

\bibitem {Latta:2009}
C.\ Latta \textit{et al.}, Nature Phys. \textbf{5}, 758
(2009).

\bibitem{Lu:2010}
C.\ -Y\ Lu \textit{et al.}, Phys.\ Rev.\ B\ \textbf{81},
035332 (2010).

\bibitem {Finley:personal}
A.\ Laucht \textit{et al.}, arXiv:cond-mat/09123685 (2009).


\bibitem{Flagg:2010}
E.\ B.\ Flagg \textit{et al.}, Phys.\ Rev.\ Lett.\ \textbf{104},
137401 (2010).

\bibitem{Xu:2007}
X.\ Xu \textit{et al.}, Phys.\ Rev.\ Lett.\ \textbf{99},
097401 (2007).

\bibitem {Koppens:2006}
F.\ H.\ L.\ Koppens \textit{et al.}, Nature (London) \textbf{442}, 766
(2006).

\bibitem{Nielsen-Chuang}
M.\ A.\ Nielsen and I.\ L.\ Chuang, \textit{Quantum Computation and Quantum Information}
(Cambridge University Press, 2000).

\bibitem{Hanson:2007}
R.\ Hanson \textit{et al.}, Rev.\ Mod.\ Phys.\ \textbf{79},
1217 (2007).

\bibitem{Larque:2009}
M.\ Larqu\'e \textit{et al.}, New J.\ Phys.\ \textbf{11},
033022 (2009).

\bibitem {Strauf:2007}
S.\ Strauf \textit{et al.}, Nature Photonics \textbf{1}, 704
(2007).

\bibitem{Englund:2009}
D.\ Englund \textit{et al.}, Optics\ Express\ \textbf{17},
18651 (2009).

\bibitem{Engel:2006}
H.\ A.\ Engel \textit{et al.}, arXiv:cond-mat/0612700 (2006)

\bibitem{Kim:2008}
D.\ Kim \textit{et al.}, Phys.\ Rev.\ Lett.\ \textbf{101},
236804 (2008).

\bibitem {Gerardot:2008}
B.\ D.\ Gerardot \textit{et al.}, Nature (London) \textbf{451}, 441
(2008).

\end{thebibliography}
\end{document}